\documentclass[onecolumn,nobibnotes,nofootinbib]{revtex4}
\usepackage{amsmath,amssymb}
\usepackage{epsfig}
\usepackage{graphicx}

\newcommand{\x}{\mathbf{x}}
\newcommand{\y}{\mathbf{y}}
\newcommand{\z}{\mathbf{z}}
\newcommand{\vecr}{\mathbf{r}}
\newcommand{\vecb}{\mathbf{b}}
\newcommand{\vk}{\mathbf{k}}

\newcommand{\ab}{\bar\alpha}

\begin{document}
\title{\bf $\gamma^*p$ cross section from the dipole model in momentum space}
\author{J. T. de Santana Amaral}
\email{thiago.amaral@ufrgs.br}
\affiliation{Instituto de F\'{\i}sica, Universidade Federal do Rio Grande do Sul, Caixa Postal 15051, 91501-970 - Porto Alegre, RS, Brazil}

\author{M. A. Betemps}
\email{marcos.betemps@ufpel.edu.br}
\affiliation{Instituto de F\'{\i}sica, Universidade Federal do Rio Grande do Sul, Caixa Postal 15051, 91501-970 - Porto Alegre, RS, Brazil}
\affiliation{Conjunto Agrot\'{e}cnico Visconde da Gra\c{c}a, Universidade Federal de Pelotas, Caixa Postal 460, 96060-290 - Pelotas, RS, Brazil}

\author{M. B. Gay Ducati}
\email{beatriz.gay@ufrgs.br}
\affiliation{Instituto de F\'{\i}sica, Universidade Federal do Rio Grande do Sul, Caixa Postal 15051, 91501-970 - Porto Alegre, RS, Brazil}

\author{G. Soyez\footnote{On leave from the fundamental theoretical physics group of the University of Li\`{e}ge}}
\email{g.soyez@ulg.ac.be}
\affiliation{LPTHE, Universit\'e Pierre et Marie Curie (Paris 6), Universit\'e Diderot (Paris 7), Tour 24-25, 5e Etage, Bo\^ite 126, 4 place Jussieu, F-75252 Paris Cedex 05, France}

\begin{abstract}
We reproduce the DIS measurements of the proton structure function at high energy from the dipole model in momentum space. To model the dipole-proton forward scattering amplitude, we use the knowledge of asymptotic solutions of the Balitsky-Kovchegov equation, describing high-energy QCD in the presence of saturation effects. We compare our results with the previous analysis in coordinate space and discuss possible extensions of our approach.
\end{abstract}

\maketitle

\section{Introduction}
One of the most intriguing problems in Quantum Chromodynamics (QCD) is the growth of the cross sections for hadronic interactions with energy. As well-known, the increase of energy causes a fast growth of the gluon density and consequently of the cross sections. At very high energies, this growth should not continue indefinitely and at some point, one has to deal with gluon recombination and multiple scattering in order to restore unitarity. This interaction between overlapping partons is called saturation and has deserved active studies over the last thirty years \cite{glr,muelqiu,ahm90,agl,jkmw97,kovmuel98}.

More generally, the large amount of work devoted to the description and understanding of perturbative QCD in the high-energy limit covers the description of saturation on the theoretical side as well as its applications to phenomenology. The theoretical contribution comes mainly from the development of non-linear QCD equations describing the evolution of scattering amplitudes towards this limit, together with the search of the solutions to those equations. The simplest of such equations is the Balitsky-Kovchegov (BK) equation \cite{bal,kov}, which corresponds to the Balitsky-Fadin-Kuraev-Lipatov (BFKL) \cite{bfkl} linear evolution equation with the addition of a non-linear term responsible for the saturation of the growth of gluon density. It has been shown \cite{mp} that the BK equation is in the equivalence class of the Fisher-Kolmogorov-Petrovsky-Piscounov (FKPP) nonlinear partial differential equation \cite{fkpp}, which admits travelling-wave solutions, translating, in terms of QCD variables, into {\em geometric scaling} as we shall explain below.

From the phenomenological side, the geometric scaling has been observed at the DESY $ep$ collider HERA, in the measurements on inclusive $\gamma^*p$ scattering \cite{gscaling}. This phenomenological feature of high-energy Deep Inelastic Scattering (DIS) is expressed as a scaling property of the virtual photon-proton cross section
\begin{equation}\label{cross_scaling}
\sigma^{\gamma^{*}p}(Q^2,Y)=\sigma^{\gamma^{*}p}\left(\frac{Q^{2}}{Q_{s}^{2}(Y)}\right),
\end{equation}
that is, the cross section depends on the scaling variable $\tau=Q^2/Q_s^2(Y)$ instead of $Q^2$ and $Q^2_s(Y)$ separately. Here $Q^2$ is the virtuality of the photon, $Y=\log(1/x)$ is the total rapidity, $x$ is the Bjorken-$x$, related to the centre-of-mass energy through $s=Q^2x$ and $Q_{s}(Y)$ is an increasing function of $Y$ called the {\em saturation scale}. The geometric scaling is actually equivalent to the formation of travelling-wave solutions for the BK equation. This is thus a remarkable consequence of saturation, which extends arbitrarily far beyond the fully saturated domain, {\em i.e.} in the dilute regime where saturation effects may seem negligible.

In this paper, we use the dipole model \cite{dipolepic} to relate the $\gamma^*p$ cross-section to the dipole-proton forward scattering. This approach has already been proven successful {\em e.g.} in \cite{gbw,iim,gb-sapeta}. Our approach here is to parametrise the dipole-proton amplitude in momentum space, where the travelling-waves have been originally derived. We shall discuss the advantages of our method and compare it with previous results in the literature later on.

The plan of this paper is as follows. In section \ref{sec:dipole}, we relate the $\gamma^*p$ cross section to the dipole-proton scattering amplitude within the dipole framework. We then discuss, in Section \ref{sec:qcd} how one can describe the dipole scattering amplitude from the properties of the BK equation. In Section \ref{sec:model} we gather all information to build the complete model for the proton structure function. The fitting procedure used to compare our model with the experimental measurements is explained in Section \ref{sec:fit} and the results of the fit are presented in Section \ref{sec:res}. We discuss the link with previous approaches in the literature and possible situations in which our work can find interesting applications in Section \ref{sec:ccl}.

\section{the dipole model}\label{sec:dipole}

\begin{figure}[h!]
\includegraphics[scale=0.9]{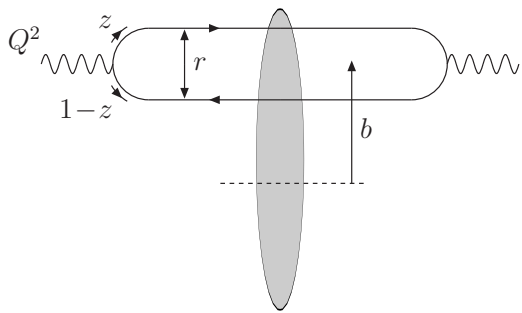}
\caption{\label{fig:dipole} Picture representing the dipole model.}
\end{figure}

We consider the collision between a virtual photon and a proton at high energy. In a frame where the photon travels fast, one can consider that it fluctuates into a $q\bar{q}$ dipole. The lifetime of this dipole being much longer than the time of interaction with the proton, one can write the cross section as a product of the wavefunction for a photon to go into a dipole times the dipole-proton cross section, as shown on figure \ref{fig:dipole}. It leads to the well-known formula
\begin{equation}
\sigma_{T,L}^{\gamma^{*}p}(Q^2,Y)=\int d^2r\int_{0}^{1}dz\,\left|\Psi_{T,L}(r,z;Q^2)\right|^{2}\sigma_{\text{dip}}^{\gamma^{*}p}(r,Y),\label{cross_section}\end{equation}
where $\sigma_{\text{dip}}^{\gamma^{*}p}(r,Y)$ is the dipole-proton cross section. The transverse and longitudinal photon wavefuctions in this expression are computable in perturbative QED. They are given by
\begin{eqnarray}
\vert \Psi_{T}(r,z;Q^2) \vert^2 
  & =& \frac{N_c\alpha_{em}}{2\pi^2}\sum_{q}e^2_q
       \left\{\left[z^2+(1-z)^2\right]\bar Q_q^2K^2_1(\bar{Q}_q r)+m^2_qK^2_0(\bar{Q}_q r)\right\}\label{eq:psit}\\
\vert \Psi_{L}(r,z;Q^2) \vert^2
  & =& \frac{N_c\alpha_{em}}{2\pi^2}\sum_qe^2_q
       \left\{4Q^2z^2(1-z)^2K^2_0(\bar{Q}_q r)\right\}\label{eq:psil}
\end{eqnarray}
where $N_c$ is the number of colours, $\bar{Q}_q^2=z(1-z)Q^2+m^2_q$, $m_q$ the mass of the quark of flavour $q$ and $K_{0,1}$ are the Mc Donald functions of rank zero and one, respectively.
This expression extends the Kovchegov formula (Eqs. (4), (5a) and (5b) in the second reference of \cite{kov}) by including the mass of the quarks.

If one treats the proton as a homogeneous disk of radius $R_p$, the dipole-proton cross section in eq.(\ref{cross_section}) is usually taken to be proportional to the dipole-proton forward scattering amplitude $T(r,Y)$ through the relation
\[
\sigma_{\text{dip}}^{\gamma^{*}p}(r,Y)=2\pi R_{p}^{2}\:T(r,Y).
\]
The proton structure function $F_2$ can be obtained from the $\gamma^*p$ cross section through the relation
\begin{eqnarray}
F_2(x,Q^2)=\frac{Q^2}{4\pi^2\alpha_{em}}\left[\sigma_T^{\gamma^*p}(x,Q^2)+\sigma_L^{\gamma^*p}(x,Q^2)\right].
\end{eqnarray}
As we shall see in the next section, the BK equation describes the high-energy evolution of the dipole-proton scattering amplitude $T$. However, the asymptotic behaviour of its solutions is naturally expressed in momentum space. Hence, we want to express the $\gamma^*p$ cross section in terms of $T(k,Y)$, the Fourier transform of $T(r,Y)$:
\begin{equation}\label{eq:tk}
T(k,Y)=\frac{1}{2\pi} \int \frac{d^2r}{r^2}\,e^{i\vk.\vecr}\,T(r,Y) = \int_0^\infty\frac{dr}{r}J_0(kr)\,T(r,Y).
\end{equation}

After a bit of algebra, we obtain the following expression relating the proton structure function to $T(k,Y)$:
\begin{eqnarray}\label{eq:final_f2}
F_{2}(x,Q^{2})=\frac{Q^{2}R_{p}^{2}N_{c}}{4\pi^{2}}\int_{0}^{\infty}\frac{dk}{k}\int_{0}^{1}dz\,|\tilde{\Psi}(k^{2},z;Q^{2})|^{2}T(k,Y),
\end{eqnarray}
where the wavefunction is now expressed in momentum space
\begin{eqnarray}\label{eq:wave_mom}
|\tilde{\Psi}(k^{2},z;Q^{2})|^{2}
  & = & \sum_q \left(\frac{4\bar{Q}_q^2}{k^2+4\bar{Q}_q^2}\right)^2 e_q^2
        \left\{ \left[z^2+(1-z)^2\right]
        \left[\frac{4(k^2+\bar{Q}_q^2)}{\sqrt{k^2(k^2+4\bar{Q}_q^2)}}\textrm{arcsinh}\left(\frac{k}{2\bar{Q}_q}\right) 
             +\frac{k^2-2\bar{Q}_q^2}{2\bar{Q}_q^2}
        \right]\right.\\
  & & \phantom{\sum_q \left(\frac{4\bar{Q}_q^2}{k^2+4\bar{Q}_q^2}\right)^2 e_q^2}
      + \left. \frac{4Q^2z^2(1-z)^2+m^2_q}{\bar{Q}_q^2}
        \left[\frac{k^2+\bar{Q}_q^2}{\bar{Q}_q^2}
             -\frac{4\bar{Q}_q^4+2\bar{Q}_q^2k^2+k^4}{\bar{Q}_q^2\sqrt{k^2(k^2+4\bar{Q}_q^2)}}
             \textrm{arcsinh}\left(\frac{k}{2\bar{Q}_q}\right)\right]
	\right\}.\nonumber 
\end{eqnarray}

Having described the DIS structure function through (\ref{eq:final_f2}), the approach used to describe
the scattering amplitude $T(k,Y)$, the forward scattering amplitude in momentum space, will be presented and discussed in the next section.

\section{Scattering amplitudes in high-energy QCD}\label{sec:qcd}

Let us now consider a fast-moving colourless $q\bar{q}$ dipole of transverse size $r=|\x-\y|$, where $\x$ and $\y$ are the coordinates of the quark and antiquark respectively, interacting with a given dense target. In the large-$N_c$ approximation ($N_c$ being the number of colours), and in the mean-field approximation, the high-energy behaviour of the dipole forward scattering amplitude $T(\x,\y;Y)$ follows the BK equation \cite{bal,kov}. In coordinate space this equation reads
\begin{equation}\label{eq:bkcoords}
\partial_Y T(\x,\y;Y)
 = \frac{\ab}{2\pi} \int d^2\mathbf{z}\frac{(\x-\y)^2}{(\x-\z)^2(\z-\y)^2}
   \left[ T(\x,\z;Y)+T(\z,\y;Y)-T(\x,\y;Y)-T(\x,\z;Y)T(\z,\y;Y) \right]
\end{equation}
where $\ab=\alpha_s N_c/\pi$, $\alpha_s$ is the strong coupling constant, considered fixed. If one neglects the dependence on the impact parameter $\vecb=(\x+\y)/2$ and integrates out the remaining angular dependence of $\vecr$, \eqref{eq:bkcoords} becomes an equation for $T(r,Y)$. The latter can be expressed in momentum space using \eqref{eq:tk}. 

One finds $T(k,Y)$ obeys the BK equation in momentum space
\begin{equation}\label{eq:bk}
\partial_Y T=\ab\chi(-\partial_L)T-\bar{\alpha}T^{2},
\end{equation}
where $\ab = \alpha_s N_c/\pi$ and
\begin{equation}\label{eq:kernel}
\chi(\gamma)=2\psi(1)-\psi(\gamma)-\psi(1-\gamma)
\end{equation}
is the characteristic function of the Balitsky-Fadin-Kuraev-Lipatov (BFKL) kernel \cite{bfkl} and we used $L=\log(k^2/k_0^2)$ with $k_0$ some fixed soft scale.

In our approach, we do not want to use directly numerical solutions of \eqref{eq:bk} as the input for the dipole scattering amplitude but rather use our knowledge of the properties of its solutions to build an analytical expression for $T(k,Y)$. As we shall explain in more details in the next section, this allows, for example, to take into account next-to-leading order corrections to the BFKL kernel, which provides a much better description of the data.

It has been shown \cite{mp} that, after the change of variables
\begin{eqnarray}
\label{changvar}
t=\ab Y,\quad x\sim \log(k^2/k^2_0),\quad u(x,t)\propto T(k,Y),
\end{eqnarray}
the BK equation reduces to the Fisher-Kolmogorov-Petrovsky-Piscounov (F-KPP) equation \cite{fkpp} for $u$ when its kernel \eqref{eq:kernel} is approximated in the saddle point approximation {\em i.e.} to second order in the derivative $\partial_L$, the so-called diffusive approximation. The F-KPP equation is a well-known equation in non-equilibrium statistical physics, whose dynamics is used to describe many reaction-diffusion systems in the mean-field approximation:
\begin{equation}\label{eq:fkpp}
\partial_{t}u(x,t)=\partial_{x}^{2}u(x,t)+u(x,t)-u^2(x,t),
\end{equation}
where $t$ and $x$ are, respectively, the time and space variables. 

This equation has been extensively studied since more than sixty years and, in particular, it is known to admit travelling waves as asymptotic solutions. This means that the solution $u(x,t)$ to equation \eqref{eq:fkpp} takes the form $u(x-v_c t)$ of a front travelling to large values of $x$ at a speed $v_c$ without deformation.

This property of the F-KPP equation is actually true if one considers the BK equation \eqref{eq:bk} with the full BFKL kernel \cite{mp}. At asymptotic rapidities, the amplitude $T(k,Y)$, instead of depending separately on $k$ and $Y$, depends only on the scaling variable $k^2/Q_s^2(Y)$, where we have introduced the {\em saturation scale} $Q_s^2(Y)=k_0^2\exp(v_c Y)$, measuring the position of the wavefront (more precisely, it is $L_s(Y)=\log(Q_s^2/k_0^2)=v_c Y$ which measures its position).

A more detailed calculation allows also for the extraction of two additional subleading corrections, resulting into the following expression for the tail of the scattering amplitude
\begin{equation}\label{eq:Ttail}
T\left(k,Y\right) \stackrel{k\gg Q_s}{\approx}
  \left(\frac{k^2}{Q_s^2(Y)}\right)^{-\gamma_c}\log\left(\frac{k^2}{Q_s^2(Y)}\right)
\exp\left[-\frac{\log^2\left(k^2/Q_s^2(Y)\right)}{2\ab\chi''(\gamma_c)Y}\right]
\end{equation}
where $\chi^{\prime\prime}$ denotes the second derivative of the BFKL kernel (\ref{eq:kernel}) with respect to $\gamma$ and the saturation scale
\begin{equation}\label{eq:qs}
Q_s^2(Y)=k_0^2\exp\left(\ab v_c Y-\frac{3}{2\gamma_c}\log(Y)
        -\frac{3}{\gamma_c^2}\sqrt{\frac{2\pi}{\ab\chi''(\gamma_c)}}\frac{1}{\sqrt{Y}}\right).
\end{equation}

The critical parameters $\gamma_c$ and $v_c$ are obtained from the knowledge of the BFKL kernel alone and correspond to the selection of the slowest possible wave:
\begin{equation}
v_c=\min_\gamma\ab\frac{\chi(\gamma)}{\gamma}=\ab\frac{\chi(\gamma_c)}{\gamma_c}=\ab\chi'(\gamma_c).
\end{equation}
For the leading-order BFKL kernel \eqref{eq:kernel}, one finds $\gamma_c=0.6275...,$ and $v_c=4.88\bar{\alpha}$.

Physically, the property of geometric scaling has deep consequences. It expresses the fact that when one moves along the saturation line, the behaviour of the scattering amplitudes remains unchanged. In addition, saturation introduces naturally the scale $Q_s$ which provides an infrared cut-off solving the infrared instability problem of the BFKL equation. The subleading corrections in \eqref{eq:Ttail} also have a very important role. The last term introduces an explicit dependence on the rapidity $Y$ and hence violates geometric scaling. However, this term can be neglected when
\[
\frac{\log^2\left(k^2/Q_s^2(Y)\right)}{2\ab\chi''(\gamma_c)Y} < 1.
\]
This means that geometric scaling is obtained for
\[
\log\left(k^2/Q_s^2(Y)\right) \lesssim \sqrt{2\chi''(\gamma_c)\ab Y},
\]
{\em i.e.}, in a window which extends $\sqrt{Y}$ above the saturation scale. It is a remarkable property that, at high-energy, the consequences of saturation are observed arbitrarily far in the tail, where $T(k,Y)$ is much smaller than 1.

\section{Description of the $\gamma^*p$ data}

The method that will be used to describe the experimental measurements of the proton structure function can be split in three steps. We shall first build a QCD-based model for the scattering amplitude, which can directly be used in \eqref{eq:final_f2} to obtain $F_2$. Then the details concerning the remaining ingredients needed to obtain $F_2$ are given and we specify the dataset. We finally present the results.

\subsection{Dipole scattering amplitude}\label{sec:model}

Expression \eqref{eq:Ttail} only gives a description of the tail of the wavefront $T(k,Y)\ll 1$ ($k\gg Q_s$). In order to complete the description, we also need expressions for $T$ around the saturation scale and at saturation. In the infrared domain, one can show, for example by computing the Fourier transform \eqref{eq:tk} of a Heaviside function $T(r)=\Theta(rQ_s-1)$, that the amplitude behaves like
\begin{equation}\label{eq:Tsat}
T\left(\frac{k}{Q_s(Y)},Y\right)\stackrel{k\ll Q_s}{=} c - \log\left(\frac{k}{Q_s(Y)}\right)
\end{equation}
where $c$ is an unfixed constant.

Those parametrisations (Equations (\ref{eq:Ttail}) and (\ref{eq:Tsat})) describe fully the asymptotic behaviour of the amplitude and we are left with the matching around the saturation scale. The easiest way to do that is probably to use \eqref{eq:Ttail} for $k>Q_s$ and \eqref{eq:Tsat} for $k<Q_s$ and match the constant $c$ to obtain a continuous distribution. However, this definition by parts would certainly introduce oscillations in the coordinate space $T(r,Y)$ which may even lead to negative amplitudes. Thus, the best way to obtain the description of the transition to the saturation region is to perform an analytic interpolation between both asymptotic behaviours.

In order to obtain an interpolation model which describes the transition from the dilute regime to the saturation one, the idea is to build the latter domain from the former. Our starting point is an expression which is monotonically decreasing with $L$ and which reproduces (up to the logarithmic factor), the amplitude for geometric scaling \eqref{eq:Ttail} 
\begin{equation}\label{eq:Tdil}
T_{\text{dil}} = \exp\left[-\gamma_c\log\left(\frac{k^2}{Q_s^2(Y)}\right)-\frac{L_{\text{red}}^2-\log^2(2)}{2\bar{\alpha}\chi''(\gamma_c)Y}\right]
\end{equation}
with
\begin{equation}\label{eq:l_red}
L_{\text{red}}=\log \left[1+\frac{k^2}{Q_s^2(Y)}\right] \qquad\text{ and }\quad Q_s^2(Y) = k_0^2\,e^{\ab v_c Y}.
\end{equation}
This result is unitarised {\em \`a la} Glauber-Mueller {\em i.e.} $T_{\text{unit}} = 1-\exp(-T_{\text{dil}})$ and we finally need to reinsert both logarithmic behaviours in the infrared and in the ultraviolet. We obtain that the following choice gives good results:
\begin{equation}\label{eq:Tmodel}
T(k,Y) = \left[\log\left(\frac{k}{Q_s}+\frac{Q_s}{k}\right)+1\right] \, \left(1-e^{-T_{\text{dil}}}\right).
\end{equation}

Equations (\ref{eq:Tdil}), (\ref{eq:l_red}) and (\ref{eq:Tmodel}) determine our model for the scattering amplitude. They provide a valid and simple interpolation between the QCD constraints \eqref{eq:Ttail} and \eqref{eq:Tsat}. They will be considered in order to describe the structure function $F_2$ given by (\ref{eq:final_f2}).

\subsection{Dataset}\label{sec:fit}

We fit all the last HERA measurements of the proton structure function from H1 \cite{h1}, ZEUS \cite{zeus}, with our analysis being restricted to the following kinematic range:
\begin{equation}
\begin{cases} x\leq 0.01, \\
0.045\leq Q^2 \leq 150\: \textrm{GeV${}^2$}.
\end{cases}
\end{equation}
The first limit comes from the fact that our approach is meant to describe the high-energy amplitudes {\em i.e.} the small $x$ behaviour. The second cut prevents to reach too high values of $Q^2$ for which DGLAP corrections need to be included properly. This gives a total amount of 279 data points. In addition, we have allowed for a 5\% renormalisation uncertainty on the H1 data. 

Concerning the parameters, we have kept $\gamma_c=0.6275$ and $\bar{\alpha}=0.2$ fixed although a particular choice for $\alpha_s$ only results in a renormalisation of the other parameters. For the convolution with the photon wavefunction, we have assumed different situations for the quarks masses: the light-quarks mass $m_q$ has been set to 50 or 140 MeV while, we have used $m_c=m_q$ or $m_c=1.3$ GeV for the charm mass.
This leaves $v_c$, $\chi_c''$, $k_0^2$ and $R_p$ as free parameters.

\subsection{Results}\label{sec:res}

\begin{figure}
\includegraphics{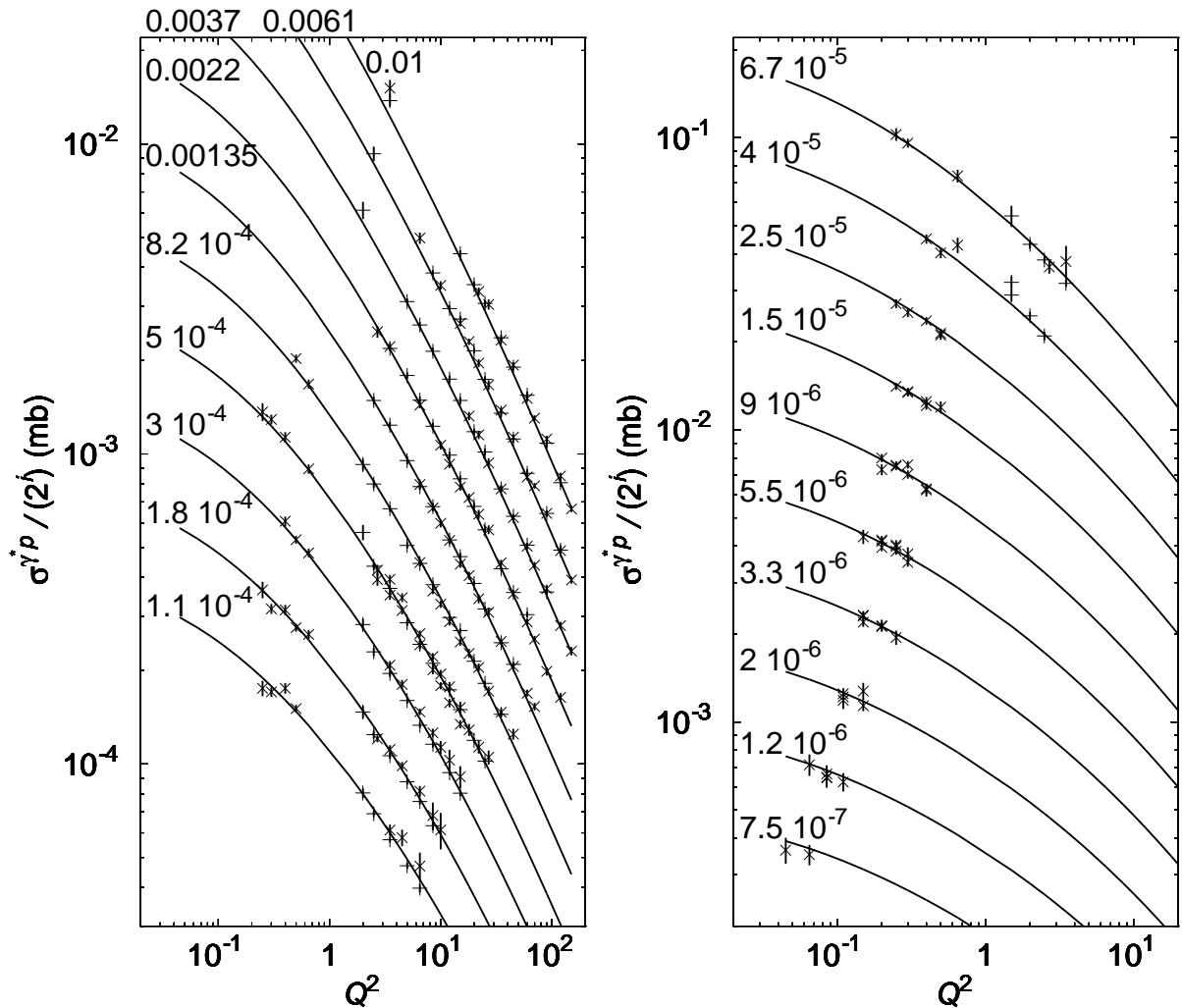}
\caption{Results for the fit to the virtual photon-proton cross section. The data points and the fit are shown as function of $Q^2$ for different values of $x$. Those $x$ values are indicated for each curve. For both plots, the successive curves have been rescaled by powers of 2 (1 to $2^{-9}$ from top to bottom) to ensure clarity. The plus corresponds to the ZEUS points, the cross to H1.}\label{fig:f2p}
\end{figure}

\begin{figure}
\includegraphics[scale=0.8]{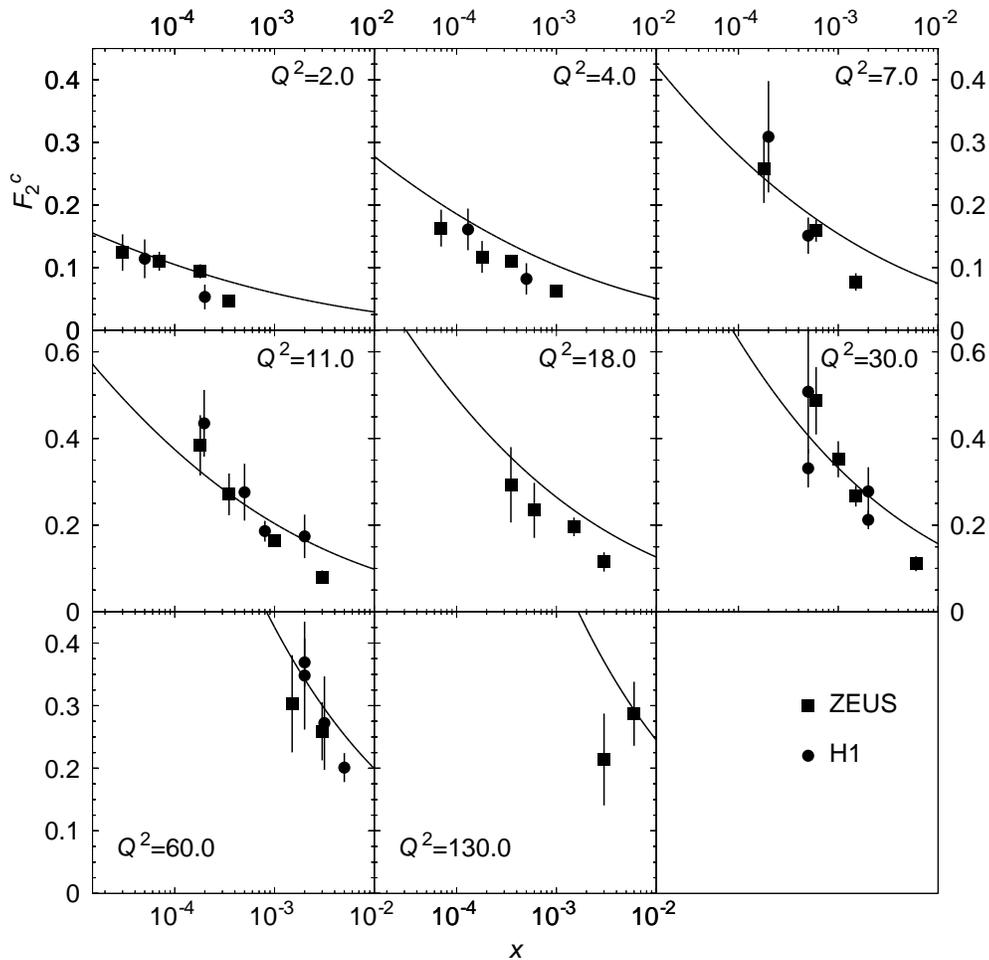}
\caption{Predictions for the H1\cite{h1c} and ZEUS\cite{zeusc} measurements of the charm structure function. The result is presented as a function of $x$ for different values of $Q^2$. For each subplot, the value of $Q^2$ is given in GeV${}^2$.}\label{fig:f2c}
\end{figure}

\begin{table}
\begin{center}
\begin{tabular}{|l||c|c|c|c||c|}
\hline 	
Masses & $k_0^2$ ($10^{-3}$ GeV${}^2$) & $v_c$ & $\chi''_c$ & $R_p$ (GeV) & $\chi^2$/nop \\
\hline
$m_q= 50$ MeV, $m_c= 50$ MeV & $3.782 \pm 0.293$ & $1.065 \pm 0.018$ & $4.691 \pm 0.221$ & $2.770 \pm 0.045$ & 0.960 \\
$m_q= 50$ MeV, $m_c=1.3$ GeV & $7.155 \pm 0.624$ & $0.965 \pm 0.017$ & $2.196 \pm 0.161$ & $3.215 \pm 0.065$ & 0.988 \\
$m_q=140$ MeV, $m_c=1.3$ GeV & $3.917 \pm 0.577$ & $0.807 \pm 0.025$ & $2.960 \pm 0.279$ & $4.142 \pm 0.167$ & 1.071 \\
\hline
\end{tabular}
\end{center} 
\caption{Results from the fit to the $F_2$ data. The values of the parameters with their respective errors are indicated, together with the $\chi^2$ per data point.}\label{table:res}
\end{table}

The parameters obtained from the fit are shown together with the $\chi^2$ per point in Table \ref{table:res}. We have also plotted the comparison with the experimental data for $F_2^p$ on figure \ref{fig:f2p} as well as the charm structure function $F_2^c$ on figure \ref{fig:f2c}. For both cases, the curve correspond to $m_q=50$ MeV and $m_c=1.3$ GeV. We verify a good agreement with the measurements of $F_2^p$ due to the small $\chi^2$ provided by the fit. Moreover, the $F_2^c$ predicted by the parametrisation is in reasonable agreement with the experimental results, which shows the robustness of the model proposed in this work.

Within our parametrisation, the saturation scale $Q_s$ corresponds to the energy-dependent scale at which the dipole scattering amplitude\footnote{Note that, when working in momentum space, the logarithmic behaviour of the amplitude in the infrared allows it to take values larger than 1. This is not a contradiction with unitarity.} is $T(k=Q_s(Y), Y)=[1+\log(2)](1-1/e)\approx 1.07$. For the fit corresponding to $m_q=50$ MeV and $m_c=1.3$ GeV, we obtain a saturation scale $Q_s = 0.206$ GeV for $x=10^{-4}$ and $Q_s = 0.257$ GeV for $x=10^{-5}$. Although these values may seem rather small, we have to emphasise that they correspond to large values for $T$. If, instead, we extract the saturation scale by requiring that $T=1/2$ when $k=Q_s$, we get $Q_s=0.296$ (resp. $Q_s=0.375$) GeV for $x=10^{-4}$ (resp. $x=10^{-5}$). These last values are still a bit smaller than the saturation scales observed in previous studies ($Q_s\approx 1$ GeV for $x\approx 10^{-5}$) and, hence, tend to confirm the tendency for the saturation scale to decrease when heavy-quark effects are taken into account.

\section{Conclusions and discussions}\label{sec:ccl}

In this work we have investigated the travelling-wave solutions of the BK equation which describe the forward scattering amplitude at high energies and tested their phenomenological implications for the virtual photon-proton scattering. We have proposed an expression for the amplitude in momentum space (see eqs. \eqref{eq:Tdil}, \eqref{eq:l_red} and \eqref{eq:Tmodel}) which interpolates between the behaviour of the dipole-proton amplitude at saturation and the travelling-wave, ultraviolet, amplitudes predicted by perturbative QCD from the BK equation. This expression was used to compute the proton structure function $F_2$ (in the framework of the dipole model) and tested against the HERA data. We obtained a good fit with light quark masses $m_{u,d,s}=50,140\,\textrm{MeV}$ and heavy charm mass $m_c=1.3$ GeV.

At this point, it is interesting to compare our results with those from previous approaches in the literature.
In the last decade some dipole models of DIS at small-$x$ have been proposed and turned out to be
successful in describing the experimental data. Among these, our comparisons will be particularly
focused on the pioneering Golec-Biernat-W\"usthoff (GBW) model \cite{gbw}, the Iancu-Itakura-Munier (IIM) model \cite{iim} and recent developments concerning the Bartels-Golec-Biernat-Kowalski (BGK) model \cite{gb-sapeta,bgk}, which consists of the GBW model improved by the incorporation of a proper gluon density evolving according to the DGLAP evolution equation. Some results from these models are shown in Table \ref{table:comp}. For each parametrisation, we indicate the values used for the (light and charm) quark masses, the number of points and the respective values of $\chi^2$ per data point obtained from the fit.

\begin{table}
\begin{center}
\begin{tabular}{|l||c|c|c|}
\hline 	
Parametrisation      & Quark masses                   & nop & $\chi^2$/nop \\
\hline
GBW \cite{gbw}       & $m_q= 140$ MeV, $m_c= 140$ MeV & 372 & $1.5$   \\
\hline
IIM \cite{iim}       & $m_q= 140$ MeV, no charm       & 156 & $0.81$  \\
\hline
BGK \cite{gb-sapeta} & $m_q= 0$ MeV, $m_c= 1.3$ GeV   & 288 & $1.06$  \\
\hline
This work            & $m_q= 50$ MeV, $m_c= 1.3$ GeV  & 279 & $0.988$ \\
\hline
\end{tabular}
\end{center} 
\caption{Comparison between the results from the parametrisations in \cite{gbw,iim,gb-sapeta} and the
the present work.}\label{table:comp}
\end{table}

First of all, it should be stressed that these models were developed in coordinate space and not in momentum space, as in the present work. Our choice is directly motivated by the analysis of the BK equation in momentum space leading to universal asymptotic results on which we heavily rely. 
Although most approaches are able to reproduce the $F_2$ data with a good $\chi^2$, our model can be differentiated from the previous ones at two levels. On the one hand, our analysis is based on the BK equation to account for unitarity effects. Thus, we expect it to be more precise than the Glauber-Mueller-like approach used in \cite{gbw} and \cite{gb-sapeta}, especially, in the small-$x$ and low $Q^2$ domain under study. On the other hand, we improved the IIM model by including massive charm. Moreover, in both GBW and IIM models, one finds some difficulties after performing their Fourier transform, that is, changing from the coordinate to momentum space. Such aspects were analysed in details in \cite{BG04}. In the case of GBW model, one obtains a Fourier transform of the dipole cross section which presents an unrealistic perturbative behaviour, while in the case of IIM it presents non-positivity values. These problems are tamed in our model, where the inverse Fourier transform (scattering amplitude in coordinate space) remains between 0 and 1. The resulting dipole cross section presents the colour transparency property, {\it i.e.} $\sigma_{dip}\sim r^2$ when $r\rightarrow 0$ and the saturation property, {\it i.e.} $\sigma_{dip}\sim \sigma_0$ at large $r$.

These features associated with the good data description and small $\chi^2$ provide that the dipole scattering amplitude proposed in this work should be a good parametrisation to investigate the properties of the observables at RHIC and LHC energies, considering the dipole approach.

\section*{Acknowledgements}
This work is partially supported by CAPES (J.T.S.A) and CNPq (M.B.G.D. and M.A.B.). G.S. is funded by the National Funds for Scientific Research (FNRS, Belgium).

\end{document}